\documentclass[twocolumn]{IEEEtran}

\usepackage{amsmath,epsfig,amssymb,verbatim,amsopn,cite}
\usepackage{amsthm}
\usepackage{balance}
\usepackage{multirow}
\usepackage[usenames,dvipsnames]{color}
\usepackage[all]{xy}  
\usepackage{url}
\usepackage{amsfonts}
\usepackage{amssymb}
\usepackage{epsfig}
\usepackage{epstopdf}
\usepackage{bm}
\usepackage{balance}
\usepackage{graphicx}
\usepackage{subcaption}
\usepackage{footnote}
\usepackage{cancel}
\usepackage{algorithm,algorithmic}

\usepackage{multicol} % update from RSMA[thien]
\usepackage{makecell} % update from RSMA[thien]
\usepackage[short]{optidef} 
\newtheorem{Remark}{Remark}

  {\proof}{\proofend}

\usepackage[margin=15.1mm,top=17.8mm,bottom=25.8mm]{geometry}%margins;for printer deficiencies

\newcommand{\qa}{{\bf a}}

\newcommand{\qd}{{\bf d}}
\newcommand{\qe}{{\bf e}}
\newcommand{\qf}{{\bf f}}
\newcommand{\qg}{{ \textbf{g} }}
\newcommand{\qh}{{ \textbf{h} }}
\newcommand{\qi}{{ \textbf{i} }}

\newcommand{\qs}{{\bf s}}

\newcommand{\qv}{{\bf v}}
\newcommand{\qw}{{\bf w}}
\newcommand{\qx}{{\bf x}}

\newcommand{\qB}{{\bf B}}

\newcommand{\qD}{{\bf D}}

\newcommand{\qF}{{\bf F}}

\newcommand{\qH}{{ \textbf{H} }}
\newcommand{\qI}{{ \textbf{I} }}

\newcommand{\qZ}{{ \textbf{Z} }}

\newcommand{\ettall}{\emph{et al.}}

\DeclareMathOperator{\K}{\mathcal{K}}

\DeclareMathOperator{\J}{\mathcal{J}}

\DeclareMathOperator{\C}{\mathbb{C}}

\DeclareMathOperator{\CN}{\mathcal{CN}}

\newcommand{\PZF}{\mathsf{PZF}}
\newcommand{\PMRT}{\mathsf{PMRT}}

\newcommand{\Ex}{\mathbb{E}}

\newcommand{\trace}{\mathrm{tr}}
\newcommand{\diag}{\mathrm{diag}}

\newcommand{\SEk}{\mathrm{SE}_{k}}
\newcommand{\SINRk}{\mathrm{SINR}_{k}}

\newcommand{\vA}{{\qv}_{\mathrm{A}}}

\newcommand{\vP}{{\qv}_{\mathrm{P}}}

\newcommand{\vnoise}{{v}^{\mathrm{noise}}_{k} }

\newcommand{\iA}{{\qi}_{\mathrm{A}}}
\newcommand{\iS}{{\qi}_{\mathrm{S}}}

\newcommand{\iP}{{\qi}_{\mathrm{P}}}
\newcommand{\iTX}{{\qi}_{\mathrm{T}} }

\newcommand{\ZA}{\qZ_{\mathrm{AA}}}
\newcommand{\ZS}{\qZ_{\mathrm{S}}}

\newcommand{\ZAP}{\qZ_{\mathrm{AP}}}
\newcommand{\ZPA}{\qZ_{\mathrm{PA}}}
\newcommand{\ZPP}{\qZ_{\mathrm{PP}}}
\newcommand{\ZTX}{\qZ_{\mathrm{TX}} }

\newcommand{\Zideal}{\bar{\qZ}_{\mathrm{TX}}}

\newcommand{\Ztune}{\qZ_{\mathrm{PL}}}

\newcommand{\Rloss}{R_{\mathrm{loss}}}

\newcommand{\Prad}{P_{\mathrm{rad}} }
\newcommand{\Pavail}{P_{\mathrm{avail}} }
\newcommand{\Pdel}{P_{\mathrm{del}} }
\newcommand{\PTX}{P_{\mathrm{T}} }
 
\newcommand{\Pin}{P_{\mathrm{in},j} }

\newcommand{\ERj}{\mathrm{ER}_j}
\newcommand{\IRk}{\mathrm{IR}_k}

\newcommand{\amk}{\qa_{u}}

\newcommand{\NRF}{N_{\mathrm{RF}}}

\newcommand{\FBB}{\qF_{\mathrm{BB}}}
\newcommand{\FBBk}{\qf_{\mathrm{BB},k}}
\newcommand{\FBBu}{\qf_{\mathrm{BB},u}}
\newcommand{\FRF}{\qF_{\mathrm{RF}}}
\newcommand{\FEM}{\qF_{\mathrm{EM}}}

\newcommand{\NLHEke}{\Phi_{j}^{\mathrm{NL}}}

\newcommand{\Heff}{\qH_{\mathrm{eff}}}

\newcommand{\HIR}{\qH_{\mathrm{IR}}}
\newcommand{\HER}{\qH_{\mathrm{ER}}}
\newcommand{\TBCD}{T_{\mathrm{CD}}}

\newcommand{\Zin}{\qZ_{\mathrm{in}}}

\newcommand{\tildeZA}{\bar{\qZ}_{\mathrm{AA}}}
\newcommand{\tildeZAP}{\bar{\qZ}_{\mathrm{AP}}}
\newcommand{\tildeZPA}{\bar{\qZ}_{\mathrm{PA}}}
\newcommand{\tildeZP}{\bar{\qZ}_{\mathrm{PP}}}

\begin{document}

\title{Tri-Hybrid SWIPT with Reconfigurable \\ Parasitic Antennas}

\author{Thien Duc Hua, Mohammadali Mohammadi, Hien Quoc Ngo, and  Michail Matthaiou\\
\small{
Centre for Wireless Innovation (CWI), Queen's University Belfast, U.K.\\
Email: \{dhua01, m.mohammadi, hien.ngo, m.matthaiou\}@qub.ac.uk 
}}\normalsize
\allowdisplaybreaks

% The paper headers
%\markboth{.}%
%{Shell \MakeLowercase{\textit{et al.}}: A Sample Article Using IEEEtran.cls for IEEE Journals}

% \IEEEpubid{0000--0000/00\$00.00~\copyright~2021 IEEE}
% Remember, if you use this you must call \IEEEpubidadjcol in the second
% column for its text to clear the IEEEpubid mark.

% \thispagestyle{empty} % Removes the page number from the first page
% \pagestyle{empty}     % Removes page numbers from all subsequent pages
\maketitle

\begin{abstract}
We propose a coupling-aware tri-hybrid precoding architecture for simultaneous wireless information and power transfer (SWIPT) using reconfigurable parasitic antennas. 
By integrating multiport network theory and information theory, we develop a third precoding layer that explicitly accounts for mutual coupling and reactive energy storage within the antenna domain. 
To efficiently optimize this architecture, we propose a low-complexity tri-hybrid solution comprising coordinate descent (CD) for parasitic reactive load optimization, a singular value decomposition (SVD)-based analog precoder, and a protective partial zero-forcing (PPZF) digital  precoder. Numerical results demonstrate that the transmit array introduces impedance mismatch that reduces the real power accepted by the antenna array. Moreover, the proposed tri-hybrid architecture significantly outperforms hybrid precoding by constructively exploiting the coupling effects, yielding substantial gains in spectral efficiency (SE) and sum non-linear harvested energy (NLHE) at a fraction of the hardware cost.

\let\thefootnote\relax\footnotetext{This work was supported by the U.K. Engineering and Physical Sciences Research Council (EPSRC) grant (EP/X04047X/2) for TITAN Telecoms Hub, and a research grant from the Department for the Economy Northern Ireland under the US-Ireland R\&D Partnership Programme. The work of T. D. Hua and M. Matthaiou was supported by the European Research Council (ERC) under the European Union’s Horizon 2020 research and innovation programme (grant agreement No. 101001331). The work of H.~Ngo was supported by the UK ISPF programme through the UKRI EPSRC (grant ID: UKRI554, led by University of East Anglia) under pilot project `BEAM-RAN' (UEA ref: R213867).}

\end{abstract}

\begin{IEEEkeywords}
Mutual coupling,  simultaneous wireless information and power transfer, tri-hybrid precoder. 
\end{IEEEkeywords}

%==============================================================================
\vspace{-0.5em}
\section{Introduction}
%==============================================================================
Multiple-input multiple-output (MIMO) and SWIPT are recognized as key enabling technologies for sixth-generation (6G) networks. By exploiting the spatial multiplexing capabilities of MIMO, SWIPT systems can simultaneously serve information receivers (IRs) and energy receivers (ERs), thereby enabling efficient use of the transmitted waveform \cite{bruno_swipt1}. 
More recently, hybrid precoding has been extensively explored as a cost-effective solution that reduces the number of required radio-frequency (RF) chains \cite{HuangTWC2024}. 
As a more advanced embodiment, \textit{tri-hybrid precoding} has been proposed as a scalable architecture that introduces a third layer of antenna-domain control, which allows very large effective apertures and additional electromagnetic (EM) degrees of freedom (DoF) \cite{heath_whitepp, heath_conf}.
Li \ettall~\cite{Litwc2026trihybrid} modeled EM beamforming as an antenna selection problem to maximize the sum rate and energy efficiency.
% Nevertheless, the physical consistent properties that directly affect the waveform radiation in practical SWIPT have been overlooked. While existing tri-hybrid studies introduce an antenna-domain precoding layer, they tend to neglect the mutual coupling effect by spacing adjacent antennas at distances larger than half a wavelength, thereby assuming uncoupled elements. Moreover, we highlight the unexplored investigation of the coupled parasitic-assisted tri-hybrid precoding concept, in which parasitic antennas are advantageous since they are driven entirely by the mutual coupling profile without requiring additional excitation power. By tuning the parasitic loads, the array's current distribution and mutual coupling profile can be reconfigure to enhance impedance matching and exploit the coupling effect for enabling super-directive beams \cite{ihsan:ISWCS, ihsan:eucap}. 
Recent studies have investigated tri-hybrid architectures with reconfigurable antennas, where antenna-domain control complements analog and digital precoding. While existing tri-hybrid studies introduce an antenna-domain precoding layer, they tend to neglect the mutual coupling effect by spacing adjacent antennas at distances larger than half a wavelength, thereby assuming uncoupled elements. Moreover, we examine coupling-aware parasitic-assisted tri-hybrid precoding for multiuser SWIPT, in which parasitic antennas are advantageous since they are driven by the mutual coupling profile without additional excitation power \cite{ihsan:ISWCS, ihsan:eucap}. Building on this, we consider multiuser SWIPT with separate IRs and ERs and incorporate the source-to-array power transfer efficiency (PTE) ratio into the evaluation of the three-layer precoder. To this end, our novel contributions are:
\vspace{-0.2em}
\begin{itemize}
    \item We investigate a coupling-aware tri-hybrid architecture for multiuser SWIPT using reconfigurable parasitic antennas. The formulation connects the antenna-domain current distribution with analog and digital precoding through a multiport circuit model. We incorporate the PTE metric into the SWIPT evaluation to measure the fraction of available source power accepted by the antenna structure. 
    % By accounting for the impedance mismatch induced by the precoder and antenna coupling, PTE enables the signal-processing design to be evaluated in terms of actual delivered power.
    \item We propose a low-complexity alternative tri-hybrid precoding solution, which comprises CD search for the parasitic loads, SVD-based analog precoding, and PPZF digital precoding.
    \item Numerical results demonstrate that the proposed scheme outperforms conventional hybrid precoding by exploiting the optimized mutual coupling profile. For a dense array setup with spacing below half a wavelength, the coupled tri-hybrid architecture yields over a two-fold improvement in the minimum SE (min SE) and a three-fold increase in the sum NLHE compared to the hybrid precoding baseline.
\end{itemize}

\textit{Notation:} 
We use bold lowercase and uppercase letters to denote vectors and matrices, respectively; the superscripts $(\cdot)^T$ and $(\cdot)^\dag$ stand for the transpose and Hermitian transpose, respectively; $\qI_{N}$ denotes an $N\times N$ identity matrix; $\boldsymbol{0}_{N \times K}$ denotes the $N \times K$ zero matrix; $\qa_j$ denotes the $j$th element of vector $\qa$; $\trace(\cdot)$ and $(\cdot)^{-1}$ denote the trace operator and matrix inverse, while $\diag(\cdot)$ constructs a diagonal matrix from its vector argument; $\Vert\cdot\Vert_2$ denotes the Euclidean vector norm and matrix spectral norm; $\|\mathbf{A}\|_\mathrm{F}^2$ denotes the squared Frobenius norm of matrix $\mathbf{A}$;  $\Im\{\cdot\}$, $\Re\{\cdot\}$, and $\angle(\cdot)$ are the imaginary component, real component, and the argument of the complex-valued argument, respectively; $\rm{atan2}(\cdot,\cdot)$ denotes the four-quadrant inverse tangent of two inputs; a circularly symmetric complex Gaussian random variable with variance $\sigma^2$ is denoted by $\mathcal{CN}(0,\sigma^2)$; Finally, $\mathbb{E}\{\cdot\}$ denotes the statistical expectation. 
%%%%%%%%%%%%%%%%%%%%%%%%%%%%%%%%%%%%%%%%%%%%%%%%%%%%%%%
%\vspace{-0.3em}
\section{System Model and Problem Formulation}~\label{sec:Sysmodel}
% \vspace{-0.3em}
%%%%%%%%%%%%%%%%%%%%%%%%%%%%%%%%%%%%%%%%%%%%%%%%%%%%%%%
We consider a SWIPT system in which $K$ IRs and $J$ ERs are simultaneously served by a coupled transmitter. The transmitter employs $A$ active antennas and $P$ passive parasitic elements, yielding a total of $N \triangleq A + P$ antennas. The transmitter is equipped with $N_{\rm RF}$ RF chains, where the number of RF chains should be lower than the number of antennas for cost efficiency \cite{heath_whitepp, CastellanosTCOM2026, Litwc2026trihybrid}. To this end, we define $\mathcal{K}$, $\mathcal{J}$, $\mathcal{N}$ as the sets of IRs, ERs, and transmit antennas, respectively. Let $\mathcal{U} \triangleq \{1, \dots, U\}$ denote the set of all receivers with $U$ indices ordered such that $u = k$ corresponds to $\mathrm{IR}_k$ ($k \in \mathcal{K}$) and $u = K + j$ to $\mathrm{ER}_j$ ($j \in \mathcal{J}$).  Moreover, $\IRk$ and $\ERj$ denote the IR $k$ and ER $j$, respectively. 
We define the data streams intended for the receivers as $\mathbf{s} \triangleq [s_1, s_2, \dots, s_{U}]^T \in \mathbb{C}^{U \times 1}$, where $U=K+J$. The first $K$ entries carry information, while the remaining $J$ entries are energy-bearing symbols. All symbols are mutually uncorrelated and have unit variance, giving $\Ex\{\mathbf{s}\mathbf{s}^\dagger\} = \qI_{U}$ \cite{material_2}.
The transmitter employs a tri-hybrid beamforming architecture with three precoding layers. 
Let $\FBB \in \C^{\NRF \times U}$ denote the digital baseband (BB) precoding matrix, which is formulated as a linear superposition of individual user streams, i.e., $\FBB \triangleq [\qf_{\mathrm{BB},1},\ldots,\qf_{\mathrm{BB},U}]$, where $\FBBu \in \C^{\NRF \times 1}$. In addition, an analog RF precoder, $\FRF \in \C^{A \times \NRF}$, maps the RF chains to the $A$ active antennas, subject to a unit-modulus constraint, i.e., $\left\vert [\FRF]_{i,j} \right\vert^{2} = 1, \forall i,j$. Finally, we denote $\FEM \in \C^{N \times A}$ as the parasitic-assisted EM precoder. To this end, the signal fed into the transmit antenna array is expressed as
%---------------------------
\begin{align}
    \qx &= \FEM \FRF \FBB \qs = \sum\nolimits_{u=1}^{U} \FEM \FRF \FBBu s_u.
\end{align}
%---------------------------
\begin{Remark}
    From the perspective of multiport network theory, the information-bearing symbols are physically realized as current excitations across the antenna ports in the frequency domain $(f)$. Thus, the current vector over all $N$ transmit antennas can be expressed as
        %---------------------------
    \begin{equation}
        \iTX(f) \in \C^{N \times 1} \triangleq \FEM(f) \FRF(f) \FBB(f) \qs(f),
    \end{equation}
    %---------------------------
    which is driven by the source current vector
    %---------------------------
    \begin{equation}
        \iS(f) \in \C^{A \times 1} \triangleq \FRF(f) \FBB(f) \qs(f).
    \end{equation}
    %---------------------------
    This formulation bridges the signal processing domain with the antenna domain, where antenna port currents generate radiated fields to convey the symbol vector $\qs(f)$. The EM-layer precoder, $\FEM$, captures the mutual coupling effects and impedance matching conditions inherent in the array. For notational simplicity, we henceforth omit the explicit frequency dependence, with all frequency-dependent quantities evaluated at the carrier frequency $f_c$.
\end{Remark}

We denote the current and excitation voltage at the active antenna by $\iA \in \C^{A \times 1}$ and $\vA \in \C^{A \times 1}$, respectively, and the induced current and terminal voltage vectors at the parasitic antennas by $\iP \in \C^{P \times 1}$ and $\vP \in \C^{P \times 1}$, respectively. Under lossless antenna conditions and the reciprocity principle in far-field (FF) communication, the voltage--current relationship of the parasitic-assisted transmit array can be expressed as \cite{nossek:circuittheory}
%--------------------
\begin{align}~\label{txArray}
    \begin{bmatrix}
    \vA  \\
    \vP  
    \end{bmatrix}
    &=\Zideal 
    \begin{bmatrix}
    \iA \\
    \iP 
    \end{bmatrix}
    =
    \begin{bmatrix}
    \tildeZA & \tildeZAP \\
    \tildeZPA & \tildeZP  
    \end{bmatrix}
    \begin{bmatrix}
    \iA \\
    \iP 
    \end{bmatrix},
\end{align}
%--------------------
where $\Zideal \in \C^{N \times N}$ denotes the antenna lossless impedance matrix; the sub-matrices $\tildeZA \in \C^{A \times A}$ and $\tildeZP \in \C^{P \times P}$ contain the self and mutual impedances within the active and parasitic antenna arrays, respectively; also, $\tildeZPA \in \C^{P \times A}$ and $\tildeZAP \in \C^{A \times P}$ denote the mutual coupling between these arrays. As reciprocity holds, we obtain $\tildeZAP = \tildeZPA^{T}$ \cite{nossek:circuittheory}. 
We denote $l_d$, $\rho_d$, $\mu_d$, and $\sigma_c$ as the dipole length, dipole radius, permeability of free-space, and conductivity of the wire, respectively. We denote $\Rloss$ as the resistance loss when  the circuits are operating. Assuming that all dipoles have identical physical properties, the dissipation loss is computed as \cite{konstantinos:superdirective:2023}
%--------------------
% \vspace{-0.1em}
\begin{align}~\label{eq:rloss}
    \Rloss 
    &\triangleq 
    \frac{
    \kappa l_d - \sin(\kappa l_d)
    }{
    4 \kappa \rho_d \sin^{2}( \kappa l_d / 2)
    }
    \sqrt{
    \frac{
    \mu_d f_c
    }{
    \pi \sigma_c
    }
    },
\end{align}
%----------
where $\omega= 2 \pi f_c$, $\kappa = 2\pi/\lambda$, and $\lambda=c/f_c$ denote the angular frequency, wavenumber, and wavelength depending on speed of light $c$ and carrier frequency $f_c$. To this end, the lossy impedance matrices are obtained as $\ZTX = \Zideal + \Rloss \qI_{N}$, $\ZA = \tildeZA + \Rloss \qI_{A}$, and $\ZPP = \tildeZP + \Rloss \qI_{P}$, while $\ZAP = \tildeZAP$ and $\ZPA = \tildeZPA$.
We denote the reconfigurable loads that manipulate the radiation pattern of the parasitic antenna array as $\Ztune \in \C^{P \times P} \triangleq \diag\big( Z_{\mathrm{R},1}, \ldots, Z_{\mathrm{R},P} \big), Z_{\mathrm{R},p}=jX_{p}$, where $X_p$ is the $p$-th reactive component.\footnote{The tunable reactances can be practically implemented using finite varactors or capacitor/inductor networks. The quality factor and switching latency can introduce additional losses, and these effects can be incorporated by replacing $Z_{R,p}=j X_p$ with $Z_{R,p}= R_{L,p}+jX_p$.} Applying the Kirchhoff's voltage law to the active antenna loop, the parasitic closed circuits and relationship \eqref{txArray}, we derive the equivalents for further derivations as $\vP = - \Ztune \iP$, $\vP = \ZPA \iA + \ZPP \iP$, and $\vA = \ZA \iA + \ZAP \iP$, and
%------------------------------------
\vspace{-0.1em}
\begin{align}
    \iP &= - \big( \Ztune + \ZPP \big)^{-1} \ZPA \iA,~\label{eq:ip}
    \\
    \iA &= \left( \ZS + \Zin \right)^{-1} \ZS \iS,~\label{eq:ia}
\end{align}
%-----------------------
where $\ZS \in \C^{A \times A} $ is the source impedance matrix seen at the active antenna input ports and $\Zin\triangleq \ZA - \ZAP\left( \ZPP + \Ztune \right)^{-1} \ZPA$. Equation \eqref{eq:ia} illustrates that the current $\mathbf{i}_A$ flowing into the antenna array is reduced because the source impedance $\ZS$ is not conjugate-matched to the array input impedance matrix $\Zin$. Subsequently, the parasitic-assisted current excitation at the transmitter is formulated as 
%--------------------
\vspace{0.2em}
\begin{equation}~\label{eq:it}
    \iTX \triangleq 
    \begin{bmatrix}
        \iA \\
        \iP
    \end{bmatrix}
    =
    \begin{bmatrix}
        \qI_{A} \\
        -(\Ztune + \ZPP)^{-1} \ZPA
    \end{bmatrix} \iA.
\end{equation}
%--------------------
% in which we assume $(\Ztune + \ZPP)$ is nonsingular for all $m$, which holds when the parasitic load reactances are chosen to avoid self-resonance with $\ZPP$.
Subsequently, the parasitic-assisted antenna-domain precoding layer can be derived as
%--------------------
\vspace{0.2em}
\begin{align}
   \FEM \triangleq 
   &\begin{bmatrix}
       \qI_{A} \\
       -(\Ztune + \ZPP)^{-1} \ZPA
   \end{bmatrix}
   \times
   \left( \ZS + \Zin \right)^{-1} \ZS.
\end{align}
%--------------------
To this end, the mismatch is inherently encoded in the precoder $\FEM$. In particular, the factor $(\ZS + \Zin)^{-1} \ZS$ maps the excitation from the baseband to the antenna ports. The current-divider matrix captures the excitation-dependent delivery of source power to the terminated array. Matching is assessed through accepted real power relative to available source power.

%\vspace{-1em}
\subsection{Channel Model, Circuit-Power, and SWIPT metrics}
We consider a uniform planar array (UPA) in the $xy$-plane with $A$ rows and $1+P/A$ columns. The $A$ active elements occupy the central column, while the $P$ parasitic elements occupy symmetric columns on either side. The adjacent-column and adjacent-row spacings are denoted by $d_x$ and $d_y$, respectively. Let $(x_n,y_n)$ denote the position of antenna $n$, with indices $1\leq n\leq A$ corresponding to active elements and $A+1\leq n\leq N$ to parasitic elements.
Under the unilateral approximation, the $k$-th and $(K+j)$-th rows of $\mathbf{Z}_{\mathrm{RT}} \in \mathbb{C}^{U \times N}$ model the downlink channels to $\mathrm{IR}_k$ and $\mathrm{ER}_j$, respectively \cite{nossek:circuittheory, SaabACSSC2020}.
Let $(x_u^{\mathrm{R}}, y_u^{\mathrm{R}})$ denote the coordinates of receiver $u$ relative to the array center. Its distance and azimuth angle of departure (AoD) are $r_u = \sqrt{(x_u^{\mathrm{R}})^2 + (y_u^{\mathrm{R}})^2}$ and $\phi_u = \operatorname{atan2}\left(y_u^{\mathrm{R}}, x_u^{\mathrm{R}} \right)$. Under the FF plane-wave approximation, the steering vector $\mathbf{a}_u \in \mathbb{C}^{N \times 1}$ has elements $[\amk]_n = e^{-j\varrho_{u,n}}$, where
\begin{align}~\label{eq:phaseterm}
    \varrho_{u,n} \triangleq \frac{2\pi f_c}{c}(x_n \cos\phi_u + y_n \sin\phi_u), \quad \forall n \in \mathcal{N}.
\end{align}
%--------------------
With propagation delay $\tau_u = r_u/c$ and Friis path loss $\gamma_u = (\lambda / (4\pi r_u))^2$,
we assume matched isotropic receive antennas and aligned polarization. To map the physical peak-current excitation to received power, we define the current-to-received-power channel coefficient (in Ohm) as
\begin{align}
    \beta_{u} = \gamma_{u} \frac{R_r}{2} D_{\mathrm{elem}}\left(\frac{\pi}{2}, \phi_u \right),
\end{align}
where $R_r$ is the lossless dipole feed-point radiation resistance and $D_{\mathrm{elem}}$ is the directivity obtained from the dipole radiation model in \cite[Sec. II]{konstantinos:superdirective:2023}. Accordingly, the line-of-sight (LoS) channels to $\IRk$ and $\ERj$ are formulated as
%------------
% \vspace{-0.5em}
\begin{align}
    \qh_k &\triangleq \sqrt{\beta_{k}} e^{-j2\pi f_c \tau_k} \qa_k, \forall k \in \K,
    \\
    \qg_j &\triangleq \sqrt{\beta_{K+j}} e^{-j 2 \pi f_c \tau_{K+j}} \qa_{K+j}, \forall j \in \J. 
\end{align}
We quantify source-to-array power delivery in the presence of mutual coupling and impedance mismatch. All source and antenna currents are represented by peak-amplitude phasors. We assume
identical uncoupled source impedances $\ZS=Z_0\qI_A$, where $Z_0$ denotes the
reference resistance of each RF source port. This specifies the source impedance but does not imply matching to the terminated antenna array: the actual active-port currents follow \eqref{eq:ia} and depend on $\Zin$. In the matched reference case $\Zin=Z_0\qI_A$, the load-current vector is $\iS/2$. Therefore, the available source power is defined as the maximum average power deliverable by the specified sources under conjugate matching, i.e.,
%----------
\vspace{-0.5em}
\begin{equation}~\label{eq:pavail}
    \Pavail 
    =
    \frac{Z_0}{8} \Ex\left\{ \Vert \iS \Vert^{2} \right\} 
    = 
    \frac{Z_0}{8} \left\Vert \FRF \FBB \right\Vert^{2}_{\mathrm{F}},
\end{equation}
where the expectation is taken over the symbol vector $\qs$. Equation \eqref{eq:pavail} defines a matched-reference source-power quantity and the actual antenna configuration is not assumed to satisfy the matched condition. Then, the average radiated and conductor-loss powers are formulated as 
%------------
% \vspace{-0.5em}
\begin{align}
    \Prad 
    &=  \frac{1}{2} \Ex \left\{ \iTX^{\dag} \Re\left\{  \Zideal \right\} \iTX \right\}
    \nonumber\\
    &= \frac{1}{2} \trace\left(
     \Re\left\{  \Zideal \right\}
     \FEM \FRF \FBB \FBB^{\dag} \FRF^{\dag} \FEM^{\dag} 
    \right),
    \\
    P_{\mathrm{loss}} 
    &=
    \frac{\Rloss}{2} \left\Vert \iTX \right\Vert^{2}
    =
    \frac{\Rloss}{2} \left\Vert \FEM \FRF \FBB \right\Vert^{2}_{F},
\end{align}
respectively. We denote $\Pdel$ as the average real power accepted by the antenna structure, which comprises both the radiated power and conductor dissipation. Then, the PTE ratio is obtained as $\mathrm{PTE} \triangleq \Pdel / \Pavail$, where $\Pdel = \Prad + P_{\mathrm{loss}}$. Hence, the PTE ratio quantifies the power transfer effectiveness from the source to the array. Mutual coupling and parasitic-load tuning modify the input impedance presented to the RF sources. Thus, a higher PTE indicates that a larger fraction of the available source power is accepted by the antenna array. Its implications on actual radiation depend additionally on conductor losses and the resulting current distribution.

The received signals at $\IRk$ and $\ERj$ are power-normalized complex amplitudes and their squared magnitudes and the noise variances have units of Watts, given by 
% \vspace{-0.9em}
%--------------------
\begin{align}~\label{eq:receivedsignal}
    y^{\mathtt{I}}_{k} &= \qh_{k}^{\dag} \FEM \FRF \FBBk s_{k}  
    \nonumber\\
    &\hspace{0em}
    + \sum\nolimits^{U}_{u=1, u \neq k}  \qh_{k}^{\dag} \FEM \FRF \FBBu s_{u} 
    + \vnoise,
    \\
    y^{\mathtt{E}}_{j} &= \qg_{j}^{\dag} \FEM \FRF \FBB \qs + {v}^{\mathrm{noise}}_{j},
\end{align}
%--------------------
where $\vnoise \sim \CN\big(0, \sigma^2_{k} \big)$ and ${v}^{\mathrm{noise}}_{j} \sim \CN\big(0, \sigma^2_{j} \big)$ denote the receiver noises at $\IRk$ and $\ERj$, respectively. 
Assuming perfect channel state information (CSI), the PPZF precoding design places the energy-directed beams in the null space of the effective IR channel.\footnote{This work deliberately adopts a LoS specialization to isolate the effects of mutual coupling, impedance mismatch, and parasitic-load optimization. Multipath fading and imperfect CSI are not prohibited by the proposed system model, but their joint robust treatment is beyond the scope of the study.} Thus, these beams do not contribute interference at the IRs. Accordingly, the signal-to-interference-plus-noise ratio (SINR) at $\IRk$ is 
\begin{equation}~\label{eq:sinrk}
    \SINRk = \frac{\left\vert \qh_{k}^{\dag} \FEM \FRF \FBBk \right\vert^2}
    {
    \sum_{u=1, u\neq k}^{K} \left\vert \qh_{k}^{\dag} \FEM \FRF \FBBu \right\vert^2 + \sigma_{k}^2
    }.
\end{equation}
%--------------------
Hence, we obtain the achievable SE of $\IRk$ as $\SEk = \log_{2}\left(1+\SINRk\right)$.
In addition, the RF input power at the energy harvesting (EH) circuit of the $\ERj$ can be obtained as
\begin{align}
    \Pin = \left\Vert \qg_j^{\dag}  \FEM \FRF \FBB \right\Vert_{2}^{2}.
\end{align}
%--------------------
Using the parameterized NLHE model in \cite{Boshkovska:CLET:2015}, the harvested output power is formulated as
%--------------------
\begin{align}
    \NLHEke &= \frac
    {\Psi^{\mathrm{NL}}_{j} - M_{j} \Omega_{j}}
    {1 - \Omega_{j}}, \text{ where }
    \Omega_{j}= \frac{1}{1+ \exp(a_{j} b_{j})},
    \nonumber\\
    \Psi^{\mathrm{NL}}_{j} &= \frac
    {M_{j}}{
    1 + \exp\left(-a_{j} (\Pin - b_{j} ) \right)
    },
\end{align}
where $M_{j}$ denotes the maximum DC output power at the circuitry's saturation, while $a_{j}$ and $b_{j}$ are circuit-related parameters~\cite{HuaTwc2026, HuaTcom2025}. Thus, the sum-NLHE metric is computed as $\Phi^{\mathrm{NL}}_{\mathrm{sum}} = \sum_{j=1}^{J} \NLHEke$.

%%%%%%%%%%%%%%%%%%%%%%%%%%%%%%%%%%%%%%
\vspace{-0.5em}
\subsection{Problem Formulation}
%%%%%%%%%%%%%%%%%%%%%%%%%%%%%%%%%%%%%
% In this section, we formulate the optimization problem that jointly maximizes the minimum SE across all IRs while satisfying the harvested energy QoS requirements at the ERs. 
% The max-min objective is determined to ensure strict user fairness across the network. It guarantees that all IRs receive seamless communication coverage, thus motivates the objective function. 
We denote the effective propagation matrix as $\qH_{\mathrm{eff}} \triangleq \qH^{\dag} \FEM \FRF \FBB \in \C^{U \times U} $, where $\qH \triangleq [\qh_1, \ldots \qh_{K}, \qg_1, \ldots, \qg_{J}] $. To obtain a tractable low-complexity design, we adopt the log-determinant of the effective channel as a surrogate objective. This objective promotes the aggregate channel gain, spatial rank, and conditioning required by the subsequent digital precoding design \cite{heath_conf, material_2}. Thus, the log-determinant is used as a channel-quality surrogate that favors strong and well-conditioned spatial modes.
% We highlight that the SE and $\Pin$ metrics are functions of $\Heff$, thus maximizing the spatial subspace and rank conditioning of $\Heff$ will lead to improved SWIPT performance \cite{heath_conf, material_2}.
Thus, we formulate the surrogate problem as
%---------------------
\vspace{0.5em}
\begin{subequations}\label{eq:originPF}
\begin{alignat}{2}
(\mathcal{P}1):&\max_{\FBB, \FRF, \FEM}     
& & \log \det  \!\left(
\mathbf{I}_{U} + \Heff \Heff^{\dag}
\right) \label{eq:P1:obj} 
\\
%----% constraint 0:
% &\hspace{0.5em}\text{s.t.} 
% &         &  \sum_{m=1}^{M} \Phi^{\mathrm{NL}}_{j} \geq \Gamma_{\thh}, \forall j \in \J,\label{eq:P1:ct1}
% \\
% %----% constraint 0: 
% &
% &         & \SEk \geq \mathcal{S}_{\thh}, \forall k \in \K,\label{eq:P1:ct2}
% \\
%----% constraint 0: 
&
&         & \Pavail \leq \PTX,  \label{eq:P1:ct3}
\\
% %----% constraint 0: 
% &
% &         & \QTX[m] \leq \Qmax, \forall m \in \M \label{eq:P1:ct4}
% \\
%----% constraint 0: 
&
&         & \left\vert [\FRF]_{i,j} \right\vert^{2} = 1, \forall i,j, \label{eq:P1:ct5}
\\
%----% constraint 0: 
&
&         & X_{p} \in [X_{\min}, X_{\max}], \forall p \label{eq:P1:ct6},
\end{alignat}
\end{subequations}
%------------------------
where 
% $\Gamma_{\thh}$ and $\mathcal{S}_{\thh}$ are the minimum thresholds for the harvested energy and SE at the receivers, hence 
% \eqref{eq:P1:ct1} and \eqref{eq:P1:ct2} are the HE and SE QoS constraints, respectively; 
\eqref{eq:P1:ct3} is the circuit-aware source power budget; \eqref{eq:P1:ct5} is the unit-modulus constraint at the analog precoder; finally, $[X_{\min}, X_{\max}]$ are the bounds for the parasitic loads. 
%%%%%%%%%%%%%%%%%%%%%%%%%%%%%%%%%%%%%%
\vspace{-0.6em}
\section{Parasitic-assisted Tri-hybrid Precoding Design}~\label{sec:solution}
%%%%%%%%%%%%%%%%%%%%%%%%%%%%%%%%%%%%%%%%%%%
The surrogate problem $(\mathcal{P}1)$ is non-convex due to the coupling between the EM-layer reactances, the analog phase shifters, and the digital precoder, in addition to the unit-modulus and hardware constraints.
To address this complexity, we propose a coupling-aware tri-hybrid optimization strategy that decouples the design into three stages: \textit{i)} EM-layer optimization via a CD algorithm, \textit{ii)} analog optimization via SVD, and \textit{iii)} PPZF baseband precoding to manage multi-user interference.

\vspace{-0.7em}
\subsection{EM-Layer Design} 
We first optimize the parasitic load vector $\qx_P \triangleq [X_1,\ldots,X_P]^T$ to maximize the log-determinant of the coupling-aware effective channel, which is denoted as $\qH_{\mathrm{eff, EM}} \triangleq \qH^{\dag} \FEM \in \C^{U \times A}$. To this end, the CD surrogate objective is formulated as
\begin{equation}
f(\qx_P)
= \log\det \!\left(
\mathbf{I}_{U} + \qH_{\mathrm{eff, EM}}(\qx_P)\qH_{\mathrm{eff, EM}}^{\dagger}(\qx_P)
\right).
\label{eq:cd_obj}
\end{equation}
% To this end, the objective $f(\qx_P)$ promotes an efficient rank-conditioned effective channel and improves the spatial multiplexing capability for the subsequent baseband precoder.
To solve \eqref{eq:cd_obj}, we adopt a CD strategy over a discrete search grid. 
Specifically, let the feasible reactance interval be $[X_{\min},X_{\max}]$, and then we define the uniform grid as
\begin{equation}
\mathcal{G} = \left\{ X_{\min} + \frac{g-1}{N_G-1}(X_{\max}-X_{\min}) \right\}_{g=1}^{N_G},
\end{equation}
where $N_G$ is the number of grid points. 
Then, the algorithm starts from a stochastic initialization
\begin{equation}
\qx_{P}^{(0)} \sim \mathcal{U}_{\mathrm{UD}}[X_{\min},X_{\max}]^P,
\end{equation}
where $\mathcal{U}_{\mathrm{UD}}[\cdot]$ denotes the continuous uniform distribution, and the superscript $P$ indicates that each initial reactance is drawn independently and uniformly from the feasible interval $[X_{\min}, X_{\max}]$.
At the $t$-th CD iteration, \(t=0,\ldots,T_{\mathrm{CD}}-1\), where $\TBCD$ is the total number of iterations, the current reactance vector is denoted by $\qx_P^{(t)}$. 
The CD algorithm updates one reactance at a time while keeping all other entries fixed. 
For the $p$-th parasitic element, we evaluate all candidate values $x \in \mathcal{G}$ by forming a trial vector as
\begin{equation}
\widetilde{\qx}_{P,p}^{(t)}(x)
=
\Big[
X_1^{(t+1)}, \ldots, X_{p-1}^{(t+1)}, x, X_{p+1}^{(t)}, \ldots, X_P^{(t)}
\Big]^T,
\end{equation}
where the first $(p-1)$ coordinates have already been updated in the current CD search, while the remaining coordinates are fixed from the previous iteration. 
The $p$-th coordinate is then selected as the grid point that maximizes the objective:
\begin{equation}
X_p^{(t+1)}
=
\arg\max_{x \in \mathcal{G}}
f\!\left(\widetilde{\qx}_{P,p}^{(t)}(x)\right).
\label{eq:cd_update}
\end{equation}
A full CD iteration is completed after searching through all $p = 1, \dots, P$, which yields the updated reactance vector $\qx_P^{(t+1)}$ for the next CD iteration.
We denote $J^{(t)} = f(\qx_P^{(t)})$ as the objective value after the $t$-th iteration. 
To ensure numerical stability and track the best solution encountered during the search, we store
\begin{equation}
J_{\rm best} = \max_{0\le \tau \le t+1} J^{(\tau)},
\end{equation}
with the corresponding optimized reactance vector $\qx_{P}^{\star}$. 
The iterative procedure terminates when the improvement between two consecutive outer iterations becomes sufficiently small, i.e.,
%--------------
\vspace{-0.5em}
\begin{equation}
\big|J^{(t+1)} - J^{(t)}\big|
\leq
\epsilon_{\mathrm{tol}}\big(1+|J^{(t)}|\big),
\end{equation}
%------------------
where $\epsilon_{\mathrm{tol}} > 0$ is the tolerance threshold, or after $\TBCD$ sweeps. 
Finally, the optimized parasitic-assisted EM-layer precoder is obtained as $\mathbf{F}_{\rm EM}^{\star}=\mathbf{F}_{\rm EM}(\qx_{P}^{\star}).$
% \begin{equation}
% \mathbf{F}_{\rm EM}^{\star}
% =
% \mathbf{F}_{\rm EM}(\qx_{P}^{\star}).
% \end{equation}
% The proposed CD procedure is attractive for the considered parasitic-assisted tri-hybrid architecture because the optimization variables are one-dimensional reactive loads, making exhaustive search over each coordinate feasible. 
% Moreover, each update only requires evaluating the objective over a finite grid while keeping the remaining coordinates fixed, which avoids the prohibitive complexity of joint multidimensional non-convex optimization.

\vspace{-0.7em}
\subsection{Analog Precoder Design}
Given the optimized antenna-domain precoder $\mathbf F_{\mathrm{EM}}^\star$, we design the analog precoder $\FRF$ to align the RF chain with the dominant spatial subspace of the effective channel. We obtain the updated effective matrix as $\mathbf H_{\mathrm{eff,EM}}^\star=\mathbf H^\dagger\mathbf F_{\mathrm{EM}}^\star$ and compute its full SVD, $\mathbf H_{\mathrm{eff,EM}}^\star=\mathbf U_H\mathbf\Sigma_H\mathbf V_H^\dagger$, where $\mathbf U_H\in\mathbb C^{U\times U}$ and $\mathbf V_H\in\mathbb C^{A\times A}$ are unitary matrices with orthonormal columns containing the left and right singular vectors, and $\mathbf\Sigma_H\in\mathbb R^{U\times A}$ contains the singular values in descending order. We select 
$\overline{\mathbf F}_{\mathrm{RF}} =\mathbf V_H[:,1:N_{\mathrm{RF}}]$,
% \begin{equation}
% \overline{\mathbf F}_{\mathrm{RF}} =\mathbf V_H[:,1:N_{\mathrm{RF}}],
% \end{equation}
where columns beyond the channel rank provide an orthonormal completion. The constant-modulus analog precoder is then $\mathbf F_{\mathrm{RF}}^\star =\exp\!\big(j\angle(\overline{\mathbf F}_{\mathrm{RF}})\big)$, with the operations applied elementwise.

\vspace{-0.61em}
\subsection{Digital Precoder Design}
%------------------------------------------
To strike a balance between communication and wireless power transfer performance, we employ PPZF beamforming at the digital layer of the design. By adopting the PPZF scheme, interference among the IRs is effectively managed, while ensuring complete spatial protection of the IRs from energy-directed interference intended for the ERs \cite{HuaTwc2026, HuaTcom2025}. Specifically, the PMRT precoders are designed to lie in the null space of the effective IR channel, inducing zero interference at each $\IRk, \forall k \in \mathcal{K}$, thereby protecting the SINR in \eqref{eq:sinrk}. Therefore, we implement the PPZF digital precoding for interference-cancellation in SWIPT downlink transmission. We note that SVD–PPZF construction requires $K < \NRF \leq \min\{U, A\}$, full row rank of the effective IR channel, and nonzero projected ER channels before PMRT normalization. Given optimized $\FEM^{\star}$ and $\FRF^{\star}$, the effective channel is further obtained as
\begin{align}~\label{eq:fbb-start}
    \qH_{\mathrm{eff,EM,RF}}^{\star}= \qH^{\dag} \FEM^{\star} \FRF^{\star}\in \C^{U \times \NRF}.
\end{align}
Following the procedure in \cite{HuaTwc2026, HuaTcom2025}, we partition the channel matrices from the transmitter to the IRs and ERs as
\begin{align}
    \HIR &\triangleq \qH_{\mathrm{eff,EM,RF}}^{\star}[1:K, :]\in \C^{K \times \NRF},
    \nonumber\\ 
    \HER  &\triangleq \qH_{\mathrm{eff,EM,RF}}^{\star}[K+1: K+J, :]\in \C^{J \times \NRF},
\end{align}
respectively. Subsequently, the baseband precoder $\FBB$ respectively implements PPZF and PMRT towards the IRs and ERs, which is formulated as
\begin{subequations}
 \begin{align}
    \tilde{\qw}_{k}^{\PZF} &\!=\! \HIR^{\dag} \Big(\! \HIR \HIR ^{\dag} \!\Big)^{-1} \!\!\!\!\qe_k^I 
    ,
    \qd^{\mathtt{I}}_{k} \!=\! \Vert \tilde{\qw}_{k}^{\PZF}  \Vert^{-1}_2 \tilde{\qw}_{k}^{\PZF},
    ~\label{eq:wipzf}\\
    \tilde{\qw}^{\PMRT}_{j} &\!=\! 
    \qB_{\mathrm{IR}} \HER^{\dag} \qe_{j}^{E},
    \qd^{\mathtt{E}}_{j} = \Vert \tilde{\qw}^{\PMRT}_{j}  \Vert^{-1}_2 \tilde{\qw}^{\PMRT}_{j},
    ~\label{eq:wemrt}
\end{align}   
\end{subequations}
where $\qe_{k}^{I}$ and $\qe_{j}^{E}$ are the $k$-th column of $\qI_{K}$ and the $j$-th column of $\qI_{J}$, respectively. The matrix $\qB_{\mathrm{IR}} \in \C^{\NRF \times \NRF}$ denotes the projection matrix onto the nullspace of $\HIR$, which is computed as
%---------------------------
\vspace{-0.7em}
\begin{align}~\label{eq:fbb-end}
  \qB_{\mathrm{IR}}  = \qI_{\NRF}  - \HIR^{\dag} \left( \HIR \HIR^{\dag} \right)^{-1}  \HIR,
\end{align}
%---------------------------
and satisfies $\HIR \qB_{\mathrm{IR}} = \boldsymbol{0}_{K \times \NRF}$. We construct the direction matrix as
%---------------------------
\vspace{-0.3em}
\begin{equation}
    \qD_{\mathrm{BB}}  \triangleq \left[ \qd^{\mathtt{I}}_{1}, \ldots, \qd^{\mathtt{I}}_{K}, \qd^{\mathtt{E}}_{1}, \ldots, \qd^{\mathtt{E}}_{J} \right]\in \C^{\NRF \times U}.
\end{equation}
% in which its columns have unit norm $\Vert  \qd^{\mathtt{I}}_{k} \Vert_{2} = \Vert  \qd^{\mathtt{I}}_{k} \Vert_{2} = 1$, and $\Vert \qD_{\mathrm{BB}} \Vert^{2}_{F} = U$. 
To satisfy the power budget, we compute the transmit digital precoder as $\FBB = \rho_{\mathrm{BB}} \qD_{\mathrm{BB}}$ and
\begin{equation}
    \rho_{\mathrm{BB}} = \sqrt{\frac{8 P_T}{Z_0 \Vert \FRF \qD_{\mathrm{BB}} \Vert^{2}_{F}} }.
\end{equation}
Thus, the unit-norm condition applies only to the beam directions, whereas the digital precoder satisfies $\Pavail = P_T$. It employs equal weighting of the normalized digital directions without additional per-stream power allocation variables.

% \textbf{Complexity analysis:} 
% In each CD iteration, all $P$ parasitic loads are sequentially updated, and $N_G$ candidate values are examined for each reactance. For a given candidate, forming $\qH_{\mathrm{eff,EM}}=\qH\FEM$, constructing $U \times U$ Gram matrix, evaluating the log-determinant require $\mathcal{O}(UNA)$, $\mathcal{O}(U^2A)$, and $\mathcal{O}(U^3)$, respectively. Thus, the CD complexity is $\mathcal{O}(T_{\mathrm{CD}} P N_G (UNA + U^2A + U^3))$. In addition, the SVD-based analog precoder requires $\mathcal{O}(UNA + UA\min\{U, A\})$, while the PPZF digital precoder requires $\mathcal{O}(U A N_{\mathrm{RF}} + N_{\mathrm{RF}}^2 K + N_{\mathrm{RF}} K^2 + K^3 + N_{\mathrm{RF}}^2 J)$.

\vspace{-0.5em}
\section{Numerical results}
We consider a transmitter equipped with $\NRF = 10$ RF chains, serving $K = 4$ IRs and $J = 4$ ERs within a  \(250\,\mathrm m\times250\,\mathrm m\) coverage region. The ERs are clustered in the vicinity of the base station, with their angular and radial coordinates drawn uniformly at random within a maximum cluster radius of $4$ m centered at the transmit array. 
The channel and antenna impedances are evaluated at $f_c=2.4$ GHz using a single-frequency approximation. The receiver noise is integrated over $B=200$ MHz. 
At the transmit UPA array, one parasitic element is placed on each side of every active element, yielding $P=2A$. The total transmit power budget is set to $\PTX = A \times P_n$, where $P_n$ denotes the per-antenna power budget \cite{material_2}. We set the source and radiation resistances as $Z_0= 50 \Omega$ and $R_r=50 \Omega$. The thermal noise power at the receivers is determined by Boltzmann's constant $k_B = 1.38 \times 10^{-23}\ \mathrm{J/K}$ and an ambient temperature of $T_A = 300\ \mathrm{K}$, thus $\sigma^{2}_u = k_B T_A B$. The logistic NLHE parameters are set to $a_j = 150$, $b_j = 0.024$, and $M_j = 0.024\,\mathrm{W}$ $\forall j \in \mathcal{J}$. Regarding the CD strategy for optimizing the parasitic loads, we set $X_{\min}=-100 \Omega$ and $X_{\max}=100 \Omega$, $\TBCD=20$, $N_G=21$. To validate the performance of the parasitic-assisted tri-hybrid architecture over 100 realizations, we evaluate the achievable min SE and sum-NLHE against the benchmarks:
\begin{itemize}
\item \textbf{Hybrid-blkdiag}: Hybrid precoder utilizing a block-diagonal analog phase-shifter network, representing an active-only array without parasitic elements.
\item \textbf{Hybrid-BD}: Hybrid precoder employing a fully connected analog phase-shifter network, offering maximum spatial DoF for the active-only array.
\item \textbf{Tri-hybrid-blkdiag}: A coupling-aware tri-hybrid scheme that integrates the third precoding layer with a partially connected block-diagonal analog precoder, restricting each RF chain to a disjoint subset of active antennas in a low-complexity manner \cite{heath_conf}.
\item \textbf{Tri-hybrid-BD (Proposed)}: The proposed coupling-aware tri-hybrid scheme featuring a fully connected analog precoder and CD-optimized parasitic elements, representing the fully connected configuration of the hardware-constrained architecture.
\end{itemize}
The hybrid baselines and tri-hybrid schemes are compared under the same active-antenna and RF-chain counts, available power budget, and scenario setup. However, the tri-hybrid reference array contains $40$ parasitic elements, giving it a different physical aperture. For block-diagonal precoding, the 20 active elements are partitioned into 10 two-element subarrays, each using the phase-projected dominant right singular vector of its local effective channel. Comparisons with the hybrid baselines quantify the benefit of the parasitic-assisted architecture under fixed active/RF resources.
\begin{table}[t]
\caption{\small Adjacent antenna spacings versus PTE ratio ($\%$)}
\label{table:spacing}
\vspace{-0.5em}
\footnotesize
\centering
\renewcommand{\arraystretch}{1}
\begin{tabular}{|c|c|c|c|c|}
\hline
\multirow{2}{*}{\textbf{Benchmarks}}
& \multicolumn{4}{c|}{\textbf{Adjacent spacing setup ($d_x=d_y=d$)}} \\
\cline{2-5}
& \textbf{$d=0.3\lambda$}
& \textbf{$d=0.4\lambda$}
& \textbf{$d=0.5\lambda$}
& \textbf{$d=0.6\lambda$} \\ \hline
\textbf{TB1}      & 30.03 & 47.50 & 51.52 & 46.42 \\ \hline
\textbf{TB2}      & 30.97 & 47.24 & 52.17 & 46.38 \\ \hline
\textbf{TB3}      & 49.71 & 60.15 & 53.40 & 48.14 \\ \hline
\textbf{Proposed} & 78.01 & 83.41 & 47.90 & 49.10 \\ \hline
\end{tabular}
\end{table}
%%%%%%%%%%%%%%%%%%%%%%%%%%%%%%%%%%%%

Table \ref{table:spacing} shows the PTE ratio versus antenna spacing $d \in \{0.3\lambda, 0.4\lambda, 0.5\lambda, 0.6\lambda\}$ under the settings $\TBCD=20$, $\epsilon_{\rm tol}=0.0001$, $A=20$, and $P_n=20$ dBm. To this end, \textbf{TB1} denotes the tri-hybrid scheme with all three precoding layers randomly initialized; \textbf{TB2} applies PPZF baseband precoding with random analog and EM layers; \textbf{TB3} implements PPZF digital and optimized EM precoders, while keeping random analog phase shifts; the \textbf{proposed} scheme jointly employs a tri-hybrid precoding solution. As the antennas are tightly arranged ($d = 0.3\lambda$), strong mutual coupling modifies the terminated input impedance and may increase source-to-array mismatch, thereby reducing accepted real power. This degrades the PTE across all schemes. At half-wavelength spacing ($d=0.5\lambda$), the coupling effect is weaker, which reduces the reconfiguration ability to compensate for impedance mismatch through parasitic tuning. \textit{The numerical results demonstrate that the coupling effect cannot be neglected since it significantly affects the SWIPT performance.} At $d = 0.4\lambda$, a favorable coupling regime is achieved where mutual coupling is sufficiently strong for the parasitic loads to reshape the radiation pattern while improving the impedance matching. This is confirmed by the \textbf{proposed} scheme achieving the highest PTE of $83.41\%$, exceeding \textbf{TB1}, \textbf{TB2}, and \textbf{TB3} by 35.91, 36.17, and 23.26 percentage points, respectively. Thus, we adopt $d_x = d_y = 0.4\lambda$ as the reference spacing for subsequent simulations.

% \subsection{Antenna Physics Validation}
% %%%%%%%%%%%%%%%%%%%%%%%%%%%%%%%%%%%%%%
% \begin{figure}[t]
% 	\centering
% 	\vspace{0em}
% 	\includegraphics[width=80mm]{Figure_GC/convergence_1col_ver3.eps}
% 	% \vspace{-1.0em}
% 	\caption{Convergence pattern of the CD algorithm ($\TBCD=20$, $d_x = d_y =0.4 \lambda$, $P_n=20$ dBm).\normalsize}
% 	% \vspace{0.5em}
% 	\label{fig:cdalg}
% \end{figure}
% %%%%%%%%%%%%%%%%%%%%%%%%%%%%%%%%%%%
% %%%%%%%%%%%%%%%%%%%%%%%%%%%%%%%%% 
% \begin{figure*}[t]
% \vspace{-0.5cm}
%     \centering
%     % figure
%     \begin{minipage}[t]{0.48\textwidth}
%         \centering
%         \includegraphics[trim=0 0cm 0cm 0cm,clip,width=1.00\textwidth]{Figure_GC/SEvPn_1col.eps}
%         \vspace{-1.7em}
%         \caption{ Min SE versus per-antenna power budget ($d_x\!=\!d_y\!=\!0.4\lambda$, $A \!=\!20$, $\TBCD\!=\!20$).}
%         \label{fig:SEvPn}
%     \end{minipage}
%     \hfill
%     \begin{minipage}[t]{0.48\textwidth}
%         \centering
%           \includegraphics[trim=0 0cm 0cm 0cm,clip,width=1.00\textwidth]{Figure_GC/NLHEvPn_1col.eps}
%         \vspace{-1.7em}
%         \caption{ Sum NLHE versus per-antenna power budget ($d_x=d_y=0.4\lambda$, $A =20$, $\TBCD=20$).}
%         \label{fig:HEvPn}
%     \end{minipage}
% \vspace{-1.5em}
% \end{figure*}
% %%%%%%%%%%%%%%%%%%%%%%%%%%%%%%%%%
%%%%%%%%%%%%%%%%%%%%%%%%%%%%%%%%%
\begin{figure*}[t]
\vspace{0.1cm}
    \centering
    % First figure
    \begin{minipage}[t]{0.32\textwidth}
        % \vspace{-4.7cm}
        \centering
        \includegraphics[trim=0 0cm 0cm 0cm,clip,width=1.03\textwidth]{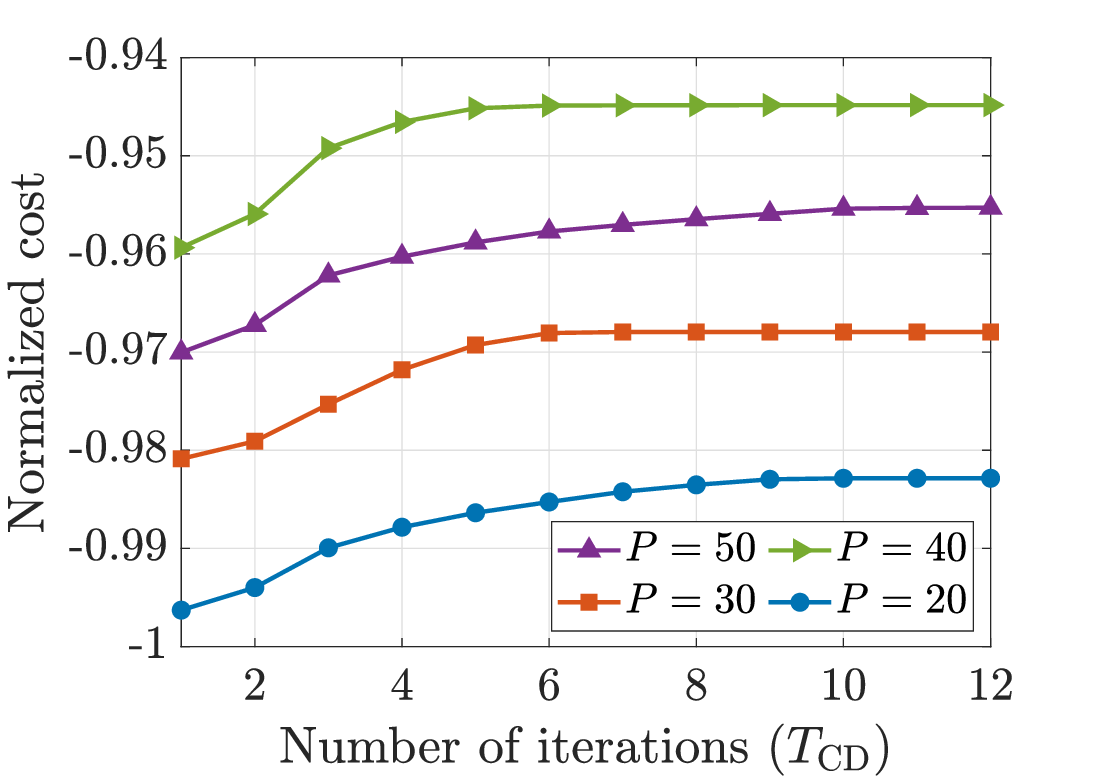}
        \vspace{-1.7em}
        \caption{\small Convergence pattern of the CD algorithm ($\TBCD=20$, $\epsilon_{\rm tol}=0.0001$, $d_x = d_y =0.4 \lambda$, $P_n=20$ dBm).\normalsize}
    \label{fig:cdalg}
    \end{minipage}
    \hfill
    % Second figure
    \begin{minipage}[t]{0.32\textwidth}
        \centering
        \includegraphics[trim=0 0cm 0cm 0cm,clip,width=1.03\textwidth]{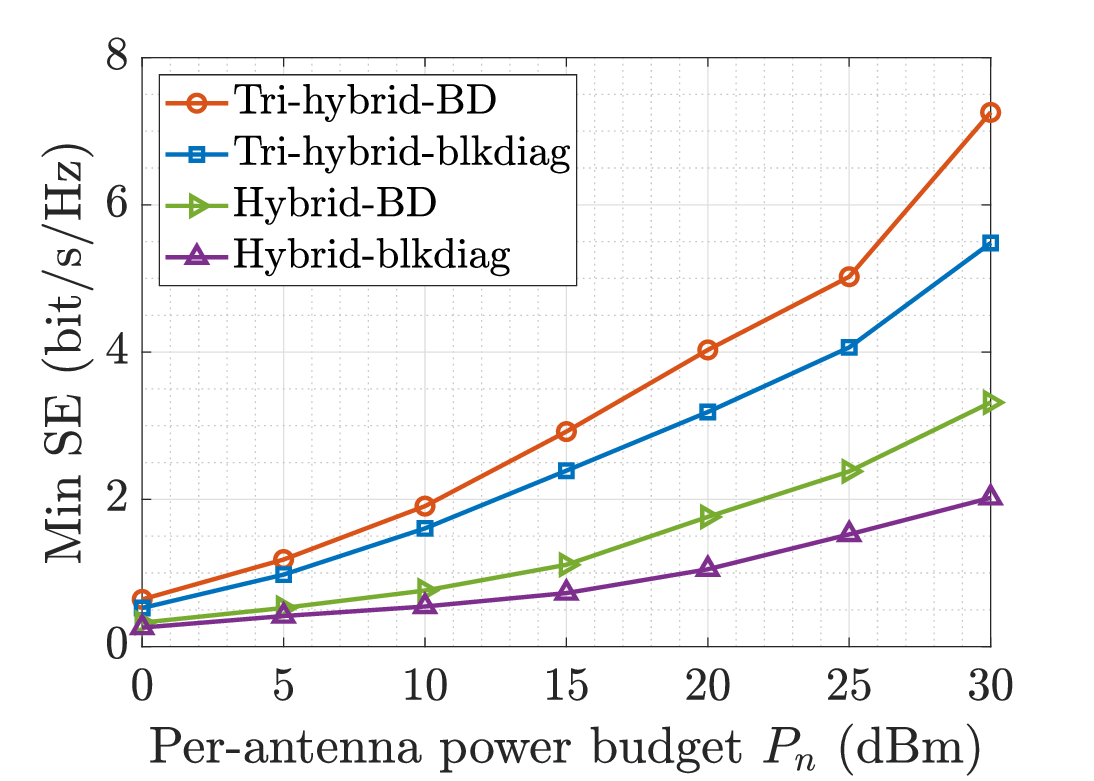}
        \vspace{-1.7em}
        \caption{\small Min SE versus the per-antenna power budget ($\TBCD=20$, $\epsilon_{\rm tol}=0.0001$, $d_x=d_y=0.4\lambda$, $A =20$).}
        \label{fig:SEvPn}
    \end{minipage}
    \hfill
    % Third figure
    \begin{minipage}[t]{0.32\textwidth}
        \centering
        \includegraphics[trim=0 0cm 0cm 0cm,clip,width=1.03\textwidth]{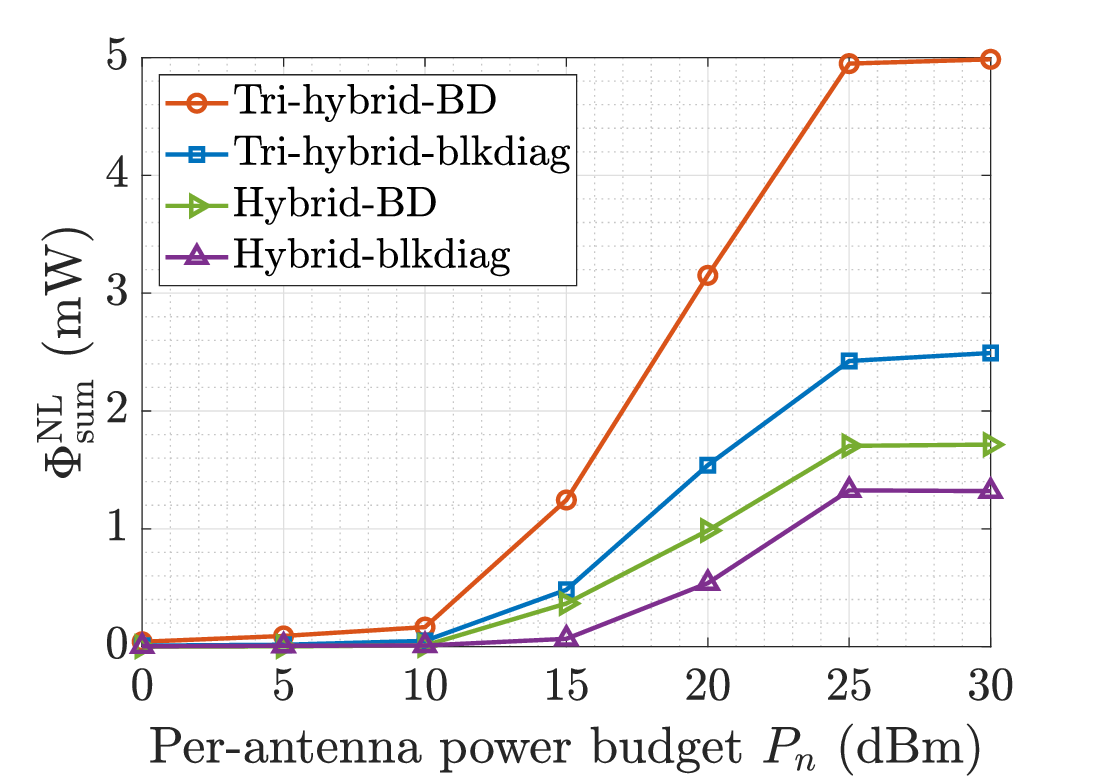}
        \vspace{-1.7em}
        \caption{\small Sum-NLHE versus the per-antenna power budget ($\TBCD=20$, $\epsilon_{\rm tol}=0.0001$, $d_x=d_y=0.4\lambda$, $A=20$).}
        \label{fig:HEvPn}
    \end{minipage}
\vspace{-1.8em}
\end{figure*}
%%%%%%%%%%%%%%%%%%%%%%%%%%%%%%%%%
Figure \ref{fig:cdalg} illustrates the convergence of the CD algorithm for varying numbers of parasitic elements $P \in \{20, 30, 40, 50 \}$, with $A = P/2$, fixed $d_x = d_y = 0.4 \lambda$, and $P_T = A \times P_n$. Thus, both the array aperture and total source-power budget vary across these configurations. The trajectories exhibit an objective improvement after 6 iterations, which confirms that the CD algorithm quickly converges within a small number of iterations regardless of the scalable array size. Notably, $P = 40$ achieves the highest objective value, while $P = 50$ yields a lower objective value despite having more antennas providing more DoF. This can be explained as follows: increasing the number of passive elements increases the mutual impedance mismatch, leading to a more complex coupling profile that makes the CD grid search prone to local optima. Thus, a physical coupling-constrained trade-off balances EM-layer exploitation for efficient beamforming. To this end, we select $A = 20$ and $P=40$ as the reference numbers of antennas for subsequent simulations.

Figure \ref{fig:SEvPn} illustrates the min SE versus the per-antenna transmit power budget $P_n$. 
The proposed tri-hybrid scheme achieves higher min SE than the active-only hybrid baseline under the same active-element and RF-chain budgets. This advantage reflects the combined effects of the parasitic aperture and load reconfiguration. In addition, the BD-analog scheme outperforms the block-diagonal counterparts. 
This can be explained that the fully connected analog architecture connects all RF chains to driven antennas, thereby yielding higher DoF to synthesize precise beamforming patterns. 

Figure \ref{fig:HEvPn} depicts the sum-NLHE versus the per-antenna transmit power budget $P_n$. In the low-power regime ($P_n \le 10$ dBm), all schemes harvest negligible NLHE since the received RF power falls below the activation sensitivity threshold of the logistic energy-harvesting circuits \cite[Remark 4]{HuaTwc2026}. As $P_n$ increases beyond $10$ dBm, the sum-NLHE rises sharply as sufficient input RF power crosses into the upper region of the logistic curve. Additionally, the \textbf{tri-hybrid-BD} scheme reaches a maximum saturated sum-NLHE of approximately $5.0$ mW, yielding a twofold improvement over the \textbf{tri-hybrid-blkdiag} scheme and significantly outperforming both the \textbf{hybrid-BD} and \textbf{hybrid-blkdiag} baselines. The SE and sum-NLHE results demonstrate the benefits of the additional parasitic aperture, coupling-aware load reconfiguration, and fully connected analog precoding under fixed active-element and RF-chain budgets.

\vspace{-0.7em}
\section{Conclusion}
This paper proposed a coupling-aware tri-hybrid precoding architecture for single-carrier SWIPT systems employing reconfigurable parasitic antennas. By integrating multiport network theory with signal-processing design, the EM-layer precoder accounts for mutual coupling and impedance mismatch. A sequential solution optimised the antenna-domain, analog, and digital precoding layers. Numerical results demonstrated substantial SE and sum-NLHE gains over conventional hybrid precoding under identical RF-chain constraints. These results support coupling-aware EM-layer design for physically consistent and energy-efficient SWIPT systems. Future work will extend the framework to wideband OFDM SWIPT and incorporate antenna Q-factor constraints addressing reactive-energy storage and frequency-dependent matching, jointly designing common parasitic loads and analog precoding with per-subcarrier digital precoding.

 \vspace{-1.2em}
% Generated by IEEEtran.bst, version: 1.14 (2015/08/26)

\vfill

\end{document}